\documentclass[pdflatex]{sn-jnl}
\usepackage{orcidlink}
\usepackage{graphicx}
\usepackage{multirow}
\usepackage{amsmath,amssymb,amsfonts}
\usepackage{amsthm}
\usepackage{mathrsfs}
\usepackage{xcolor}
\usepackage{textcomp}
\usepackage{manyfoot}
\usepackage{booktabs}
\usepackage{algorithm}
\usepackage{algorithmicx}
\usepackage{algpseudocode}
\usepackage{listings}
\begin{document}
\title[Article Title]{Review of the EU ETS Literature: A Bibliometric Perspective}
\author*[1]{\fnm{Cristiano} \sur{Salvagnin}\orcidlink{0000-0002-3348-0376}}\email{c.salvagnin@unibs.it}
\affil*[1]{\orgdiv{Department of Economics and Management}, \orgname{University of Brescia}, \orgaddress{\street{Via San Faustino 74/B}, \city{Brescia}, \postcode{25122}, \country{Italy}}}
\abstract{This study conducts a bibliometric review of scientific literature on the European Union Emissions Trading System (EU ETS) from 2004 to 2024, using research articles from the Scopus database. Using the Bibliometrix R package, we analyze publication trends, key themes, influential authors, and prominent journals related to the EU ETS. Our results indicate a notable increase in research activity over the past two decades, particularly during significant policy changes and economic events affecting carbon markets. Key research focuses include carbon pricing, market volatility, and economic impacts, highlighting a shift toward financial analysis and policy implications. Thematic mapping shows cap-and-trade systems, and carbon leakage as central topics linking various research areas. Additionally, we observe key areas where further research could be beneficial, such as expanding non-parametric methodologies, deepening the exploration of macroeconomic factors, and enhancing the examination of financial market connections. Moreover, we highlight recent and innovative papers that contribute new insights, showcasing emerging trends and cutting-edge approaches within the field. This review provides insights for researchers and policymakers, highlighting the evolving landscape of EU ETS research and its relevance to global climate strategies.}
\renewcommand{\thefootnote}{}
\footnotetext{\noindent We extend our sincere thanks to Aldo Glielmo (Bank of Italy), Maria Elena De Giuli (University of Pavia), and Antonietta Mira (Università della Svizzera Italiana) for their careful reading and the invaluable insights they shared.}
\renewcommand{\thefootnote}{\arabic{footnote}}
\maketitle
%
\section{Introduction}
\label{sec:introduction}
    The European Union presented the EU Emission Trading Scheme (EU ETS) in 2005, which was the world's first and largest international ETS, covering all the member states of Europe. Initially, the EU ETS scheme covered only $\text{CO}_2$ from manufacturing and power generation, with aviation being included in 2012, \cite{euETS2005}. The European efforts in developing and applying the continental cap-and-trade system further propelled the growth and adoption of the emissions trading scheme instrument as an effective tool to fight climate change globally. Similar to the European efforts, in 2006 the Northeast U.S. Regional Greenhouse Gas Initiative (RGGI) was established, targeting $\text{CO}_2$ emissions from the power sector in several U.S. States, \cite{rggi2006}.
        \begin{figure}[h]
        \centering
            \includegraphics[width=0.8\textwidth]{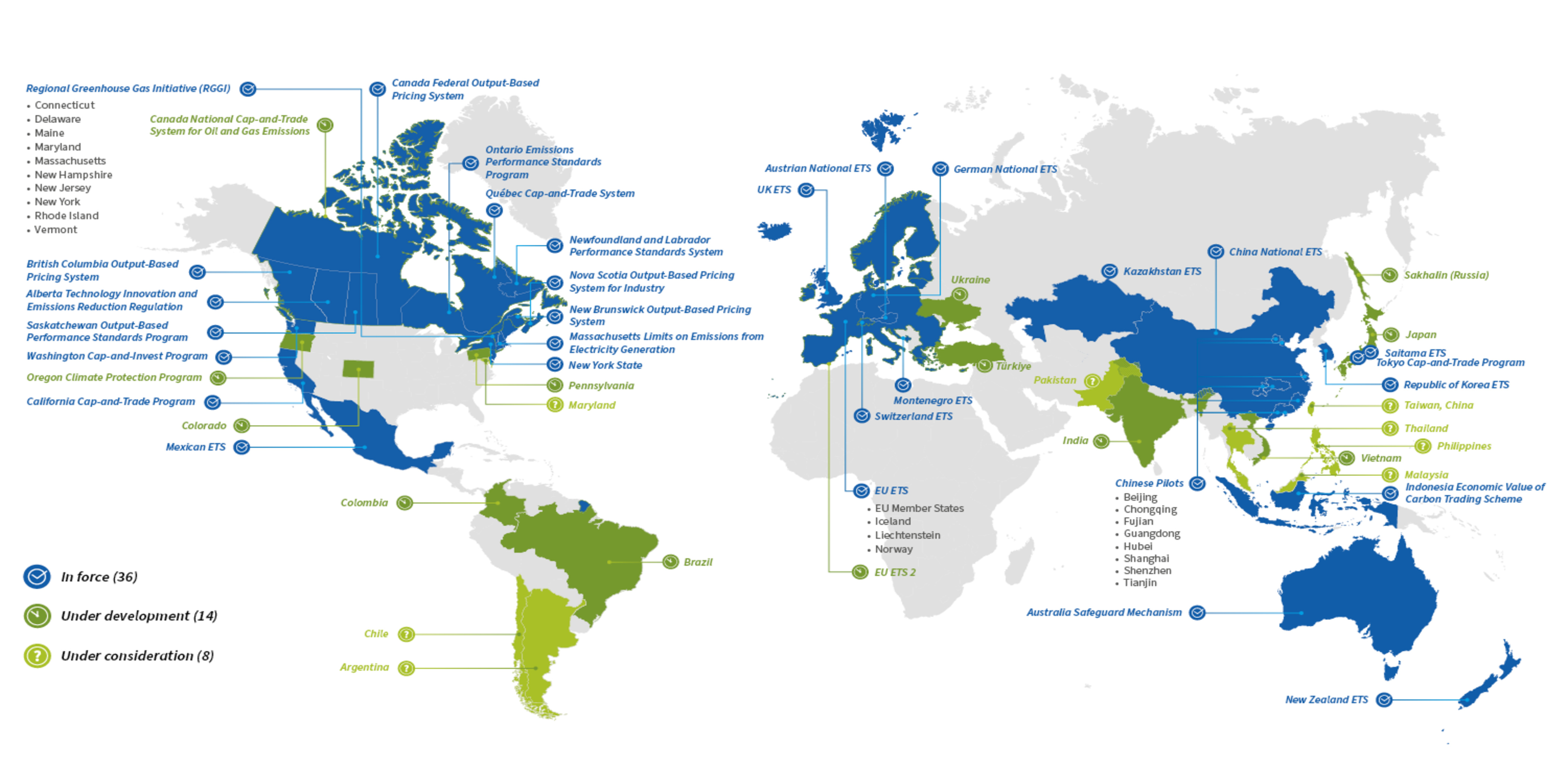}
            \caption{\textbf{ICAP World ETS Map.} This figure showcases where active and emerging emissions trading systems (ETS) are located around the globe. It highlights the increasing popularity of cap-and-trade mechanisms as countries and regions work together to tackle climate change. This image is sourced from the ICAP Carbon Action Partnership website, \textit{Copyright © 2024 International Carbon Action Partnership, accessed 10.07.2024, \href{https://icapcarbonaction.com/en/ets}{ICAP ETS Map website}.}}
        \label{fig:icap}
        \end{figure}
    In 2008, New Zealand established the NZ ETS scheme, aimed to cover all sectors of economy and with particular attention to forestry, \cite{newZealandETS2021}. In 2013, California state in U.S. launched a cap-and-trade (CaT) program, covering power generation, industry and transportation, \cite{californiaCapTrade2013}. South Korea enforce its national ETS system in 2015, becoming the first major Asian country to ever implement a allowance-based market on greenhouse gasses, \cite{southKoreaETS2015}. 
    Another milestone in the ETS adoption history is signed by the start of China's pilot programs across several provinces as Guangdong and Beijing in the late 2010, \cite{chinaETS2021}. This pilot programs served as experimental work for the lunch of the China's national ETS, one of the biggest ETS markets in the world by volume of emission coverage. 
    Finally, in 2020, a connection among EU ETS and Swiss ETS markets develop, aimed to control for cross-border trading of allowances and increasing market stability and liquidity, \cite{swissEUETS2020}. Currently, several countries, as Mexico and Canada are actively working of on the connection with American ETS markets as the California CaT program and RGGI.
    Looking ahead, more and more countries are expected to adopt or expand their ETS systems, further contributing to global emission reduction in an economic sustainable manner.
%
    \subsection{The EU ETS}
    \label{ssec:EUETS}
        The EU ETS market has been enforced since 2005, operating on a cap-and-trade principle. A limit, known as a cap, is set on the total emission allowed, which is reduced each year by the Linear Reduction Factor (LRF) percentage, \cite{EuropeanCommission2021}. The cap is divided into tradable emission permits, known as allowances. Each of these allowances allow to emit 1 tonne of $\text{CO}_2$, hence, market participants must hold a number of allowances equivalent to their emissions at the end of each compliance period, usually a calendar year, otherwise heavy fee are imposed.
        Since allowances are a tradable asset, these titles can be traded among the market participants creating a structured financial market, over-viewed by different entities as the European Commission, European Environment Agency (EEA), Member States and third-party authorities. 
        The EU ETS market is articulated in phases, each with its own duration. Phase 1, pilot phase, started in 2005 and ended in 2007. In this phase, the allocation of the allowances was mostly based on historical emissions and a significant amount of permits were freely allocated to help market participant understand the new instrument. Notably, in this phase, we observe low carbon prices due to an over allocation of permits, \cite{betz2010emissions}.
        The second phase took place from 2008 to 2012. In this phase, the EU ETS reformed and aligned with the Kyoto Protocol's commitment period and goals. This reform lowered the cap and significantly restricted the allocation of free permits, encouraging the auctioning of allowances. Despite these efforts, the economic downturn of the 2008 global financial crisis led to emission reduction causing a surplus of allowances and continued low carbon prices.
        The third phase, enforce from 2013 to 2020, brought more substantial reforms. A single EU-wide cap on emission was introduced for the first time, and the Market Stability Reserve (MSR) was implemented to face the surplus of allowances experienced in phase one and two, \cite{eudirective}. The MSR mechanism aims to prevent excessive volatility in carbon prices by automatically adjusts the supply of allowances. When the number of excess allowances exceeds a certain level, a portion of those allowances is placed into the reserve. Conversely, when the surplus is low, allowances can be released back into the market.
        Currently, the phase 4 of the EU ETS market is enforce, with an active period spanning from 2021 to 2030. Among the key strategies implemented in this phase we observe a further reduction in the cap of emissions following an accelerate annual reduction rate, to achieved the European Green Deal goals, and an expansion of the ETS to cover additional sector as maritime transport. A key goal of this phase is to align the ETS with the EU's commitment to achieving net-zero emissions by 2050.
        New developments and efforts in fighting climate change are represented by the EU ETS 2. This new market, parallel to the EU ETS, includes additional coverage to new industry sectors as buildings and road transport. This expansion is one of the instruments of the European Green Deal, which aims to reduce greenhouse gas emissions by 42\% by 2030 compared to 2005 levels, \cite{europeanGreenDeal2019}.
%
    \subsection{Literature contributions}
    \label{ssec:literature}
    This section is devoted to highlighting the key scientific contributions that outline the landscape of the EU Emissions Trading Scheme.
    In \cite{chevallier2013carbon}, the author provides a comprehensive literature-review of economics and econometrics studies regarding EU ETS, focusing on the second phase of the market. The review summarizes findings from studies on $\text{CO}_2$ price mechanism, highlighting the importance of policy measures, market fundamentals and macroeconomic factors and tendencies. Furthermore, the author point-out the gap for further investigation into how carbon prices adjust during global recession to enhance forecasting techniques and policy-support instruments for stakeholders.
    \cite{healy2015review}, found additional important contributions. In this review the focus is on the performance of the EU ETS which is evaluated under five criteria: effectiveness, efficiency, coherence, EU added value and relevance. The scope of this work covers $250$ publications on significant issues as allowances over-allocation and investment leakage. The study identifies 14 research gaps, which have been categorized into 3 main thematic areas: methodological, data and coverage gaps, stressing the need for standardized baselines, improve the data collections on investment leakage and broader coverage of new trend topics.
    More on the allowances allocation, price mechanisms and interaction with other environmental policies is presented in \cite{Zhang20101804}. In this work, key concerns in the over-allocation of allowances in phase 1, which affected carbon prices, are addressed. Moreover, the author investigates the scientific production of pricing complexity and the influence of the EU ETS allowance prices by energy costs and weather conditions. In particular, the author found significant methodological research gaps in EU ETS price mechanisms and equilibrium, suggesting future research on how rising carbon prices could affect energy companies and their emissions.
    In \cite{ji2019research}, the authors provide a bibliometric analysis of research on carbon pricing within EU ETS market, addressing several key insights and gaps. The field has seen rapid growth, with China and the USA contributing the most, and the Chinese Academy of Sciences emerging as a leading institution. Energy Policy is identified as the most productive journal in this area. Moreover, four major research areas of interest are identified: carbon price dynamics, the design of price mechanisms, carbon price policies analysis, and the effects of carbon pricing. Scholars have paid particular attention to the volatility of the market, also facing the identification of the factors influencing it. The bibliometric analysis also categorizes the factors that affect carbon pricing into five main areas: policy, emissions levels, energy prices, weather and climate, and financial markets. Key issues identified include the oversupply of allowances, inter-temporal trading, and the structure of the energy sector. Furthermore, the author emphasizes the need to explore the causes of persistently low carbon prices, conduct more quantitative analyses of influencing factors, and improve market mechanisms to enhance the effectiveness of carbon trading systems.
    Another important bibliometric contribution is from \cite{wang2018review}. The authors analyze major researches on carbon emission reduction market mechanisms. Research spans multiple disciplines, including environmental science and economics, with Energy Policy and Journal of Environmental Economics and Management being major journals. Core researchers like Pizer W. and Nordhaus W. have been highly influential in the scientific community. Hence, major research topics include decision-making under uncertainty, comparisons of carbon taxes and trading systems, and designing practical carbon market mechanisms. Notably, international collaboration in this area remains limited.
    Besides, the work from \cite{hu2022scientometric}, present a detailed review of European carbon market, focusing on its origins, core mechanisms, key research areas, and future directions. The authors argue that ETS framework is essential for tackling climate change and reducing emissions cost-effectively. A significant challenge is how carbon allowances are allocated, a topic that has yielded considerable debate among the scientific community. While free allocation based on historical emissions is suitable for transitional phases as phase 1 and 2 mostly, future phases and systems will likely shift to methods like bench-marking and auctioning.
    The literature on the EU ETS’s impact on low-carbon technology is extensive but incomplete. \cite{Teixidó2019} reviews various empirical studies on the effects of the EU ETS on innovation and low-carbon technology adoption in firms. It identifies 22 studies divided into econometric analyses and qualitative research. The econometric studies generally find that the EU ETS has positively influenced low-carbon patenting and R\&D investments among regulated firms, although some report no significant differences in emission intensity. Qualitative studies, indicate that while the EU ETS has had some impact, factors like energy costs often play a more significant role in driving technological change. Overall, the findings suggest that the EU ETS has led to modest advancements in low-carbon innovation and that the regulatory framework significantly influences firms' investment decisions.
    Research on the EU ETS has revealed limitations in the methodologies used to analyze the effects of free allocation and their causal connection with economic downturns. As noted by \cite{Laing2014509}, the investigation into the over-allocation of allowances indicates a reduced impact on emissions during such downturns. While effective monitoring and favorable carbon pricing have led to some reductions, ongoing research and new data are essential for refining our understanding and informing the design of future systems.
%
    \subsection{Goals}
    \label{ssec:goals}
        The primary goal of this literature review is to provide a bibliometric analysis of scientific research on the EU ETS (Emissions Trading System) market from 2004 to 2024. By analyzing articles indexed in the Scopus database, we aim to identify significant bibliometric trends, track the evolution of academic interest, and uncover key areas of research development in this field.
        One of the key objectives is to assess the volume of research on the EU ETS, offering insights into how the academic focus has evolved over time. This includes identifying the most prolific authors, research institutions, and countries contributing to the literature.
        Additionally, the review aims to conduct a citation analysis to identify the most influential articles and contributions in the field, helping to highlight critical works that have shaped academic and policy discussions on the EU ETS.
        Another major objective is to explore the evolution of research themes and topics related to the European carbon market. This includes examining the central themes, identifying emerging trends, and uncovering areas of consensus and debate within the literature.
        We also aim to analyze the co-authorship network to uncover collaboration patterns, central authors, and key research clusters. Finally, we will create a thematic map to identify niche themes, motor themes, basic themes, and emerging or declining topics, offering insights into the future direction of the research.
        Overall, the goal of this review is to identify research gaps in the EU ETS literature and provide guidance for future studies, with the aim of enhancing the understanding and effectiveness of emissions trading systems in achieving climate policy objectives.
%
    \subsection{Structure of the work}
    \label{ssec:structure}
        The remainder of this work is organized as follows.
        Section \ref{sec:methodology} describes the data collection process, including document selection criteria and the bibliometric tools and techniques used for analysis.
        Section \ref{sec:descriptive_analysis} offers an overview of publication trends, the geographic distribution of research, venue analysis, and most prolific authors, providing a quantitative assessment of research output and its evolution over time.
        Section \ref{sec:document_analysis} delves into the most cited documents, word analysis, and trending topics, offering insights into the most influential works and key themes in the literature.
        Section \ref{sec:conceptualstructure} maps the thematic landscape of the research field by elucidating major research themes and their interconnections through network analysis and thematic examination.
        Section \ref{sec:discussion} synthesizes the findings, identifying research gaps and suggesting areas for future research. It also highlights some recent and innovative papers, providing insights into emerging trends within the EU ETS literature.
        Finally, section \ref{sec:conclusions} summarizes the main contributions of the review, and provides a closing perspective on the significance of the findings for advancing the understanding of the EU ETS market.
%
\section{Methodology}
\label{sec:methodology}
%
    \subsection{Data collection and selection criteria}
    \label{ssec:data}
    We source data from the Scopus database due to its extensive coverage of peer-reviewed literature and robust citation tracking features.
\begin{figure}[h] 
    \centering
    \includegraphics[width=0.8\textwidth]{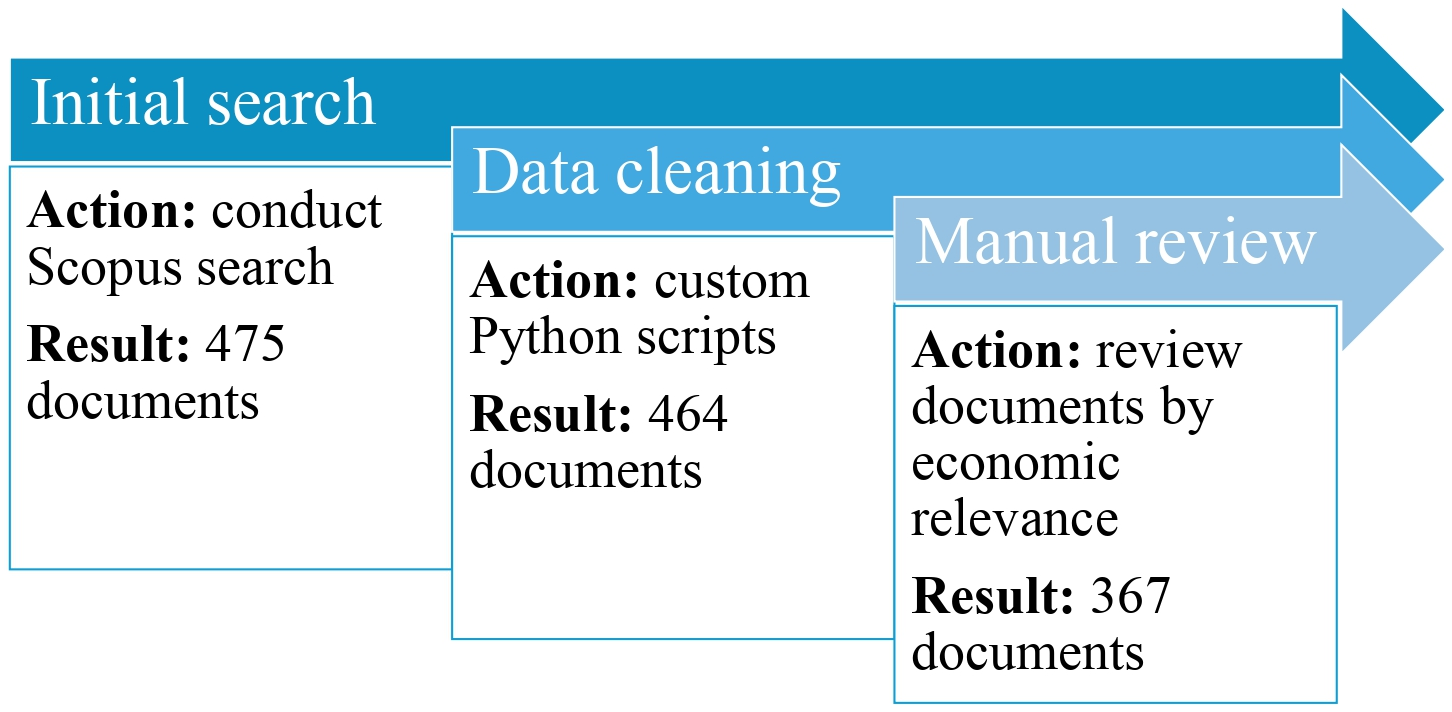}
    \caption{\textbf{Data processing workflow.} This diagram illustrates the stages of the research document selection process.}
    \label{fig:info}
\end{figure}
    The documents collected cover the period from 2004 to 2024, allowing for a detailed analysis over nearly two decades.
    We utilized predefined search terms related to the EU ETS, including \textit{European Union Emissions Trading System}, \textit{EU ETS}, \textit{emissions trading}, \textit{carbon trading}, and \textit{carbon pricing}. Boolean operators (AND, OR) helped refine our search results to ensure a broad yet targeted collection of relevant literature.
    The initial Scopus search yielded 475 documents, which were exported in BibTeX format. We improved our initial data cleaning by creating a custom Python script that automatically checked for duplicates, errors, and encoding problems. This process helped us reduce our dataset to 464 documents. Moreover, a manual review guaranteed that only papers specifically related to EU ETS market from an economic or financial perspective were considered, leading to a final selection of 367 papers, figure \ref{fig:info}.
    The relevance and quality of the literature included is ensured by considering only, peer-to-peer journal articles classified as \textit{articles}. We excluded articles from specific subject areas to retain a clear focus on the financial and economic aspects of the EU ETS market, including Chemistry, Materials Science, Medicine, Physics and Astronomy, Multidisciplinary fields, Arts and Humanities, Biochemistry, Genetics and Molecular Biology, Pharmacy, Psychology, Neurosciences, Agriculture and Biological Sciences, Civil and Structural Engineering, Engineering, Sociology and Political Science, Earth and Planetary Sciences, Decision Sciences, Mathematics, and Computer Science.
    Additionally, only articles written in English were included to maintain accessibility and comprehensibility.
%
    \subsection{Bibliometric Tools and Techniques}
    \label{ssec:tools}
        The \cite{bibliometrix} Bibliometrix R package was used for performing the bibliometric analysis.
        The analysis began with descriptive statistics to evaluate the distribution of publications over time, identifying trends in research activity and the growth rate of research articles in the EU ETS field.
        Venue analysis was performed to examine the most important journals where economic and financial EU ETS research was published. Highlighting leading journals based on publication volume and citation impact.
        Moreover, the citation analysis explore the most cited articles, influential authors, and key institutions, thereby providing insights into the impact and scope of specific studies and researchers.
        Co-authorship and collaboration networks were mapped to explore patterns of cooperation among researchers and institutions. This network analysis detected significant research networks.
        Additionally, keyword analysis was conducted to identify prevalent research themes and topics within the EU ETS literature. By analyzing keyword co-occurrence, we clustered keywords to reveal thematic trends and emerging research areas.
        Finally, the conceptual structure of the EU ETS research field was mapped using co-word analysis, visualizing the relationships between different research themes and identifying major research clusters.
%
\section{Descriptive analysis}
\label{sec:descriptive_analysis}
    \begin{table}[h]
    \centering
    \begin{tabular}{ll}
    \toprule
    \textbf{Description} & \textbf{Results} \\
    \midrule
        Timespan & 2004:2024 \\
        Sources & 85 \\
        Documents & 367 \\ 
        Authors & 731 \\
        Co-Authors per Doc & 2.67 \\
        Annual Growth Rate (\%) & 12.99 \\
        Document Average Age & 7.66 \\ 
        Average citations per doc & 42.47 \\
    \bottomrule
    \end{tabular}
    \caption{\textbf{Descriptive statistics.} This table presents the key characteristics of the dataset used for the bibliometric analysis.}
    \label{table:main_info}
    \end{table}
    This dataset features 731 authors, with an average of 2.67 co-authors per document. The annual growth rate is 12.99\%. Moreover, the average document age is 7.66 years, and each document has an average of 42.47 citations, table \ref{table:main_info}.
%
    \subsection{Publication trends}
    \label{ssec:publication_trends}
\begin{figure}[h]
        \centering
            \includegraphics[width=0.8\textwidth]{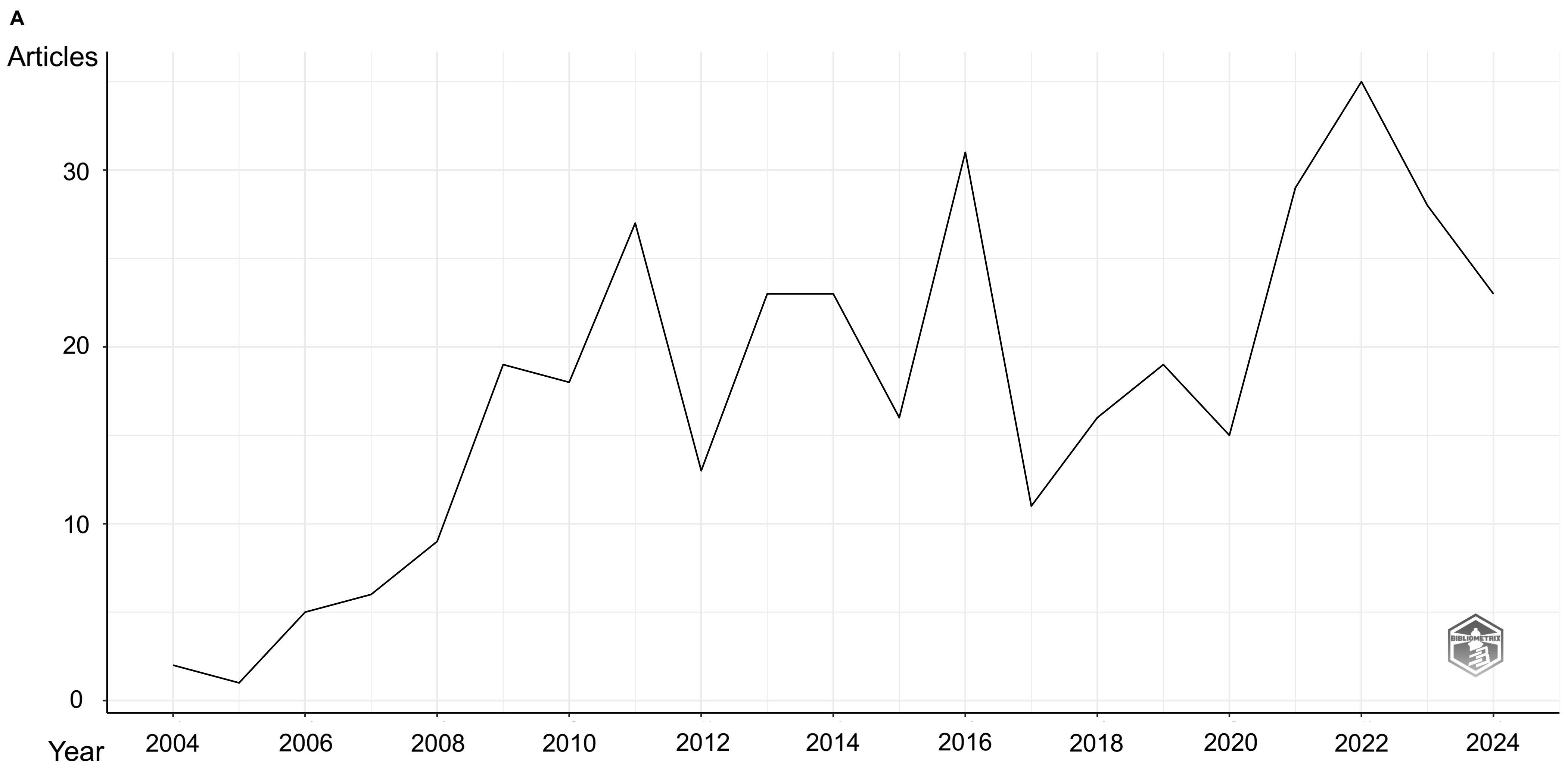}
            \includegraphics[width=0.8\textwidth]{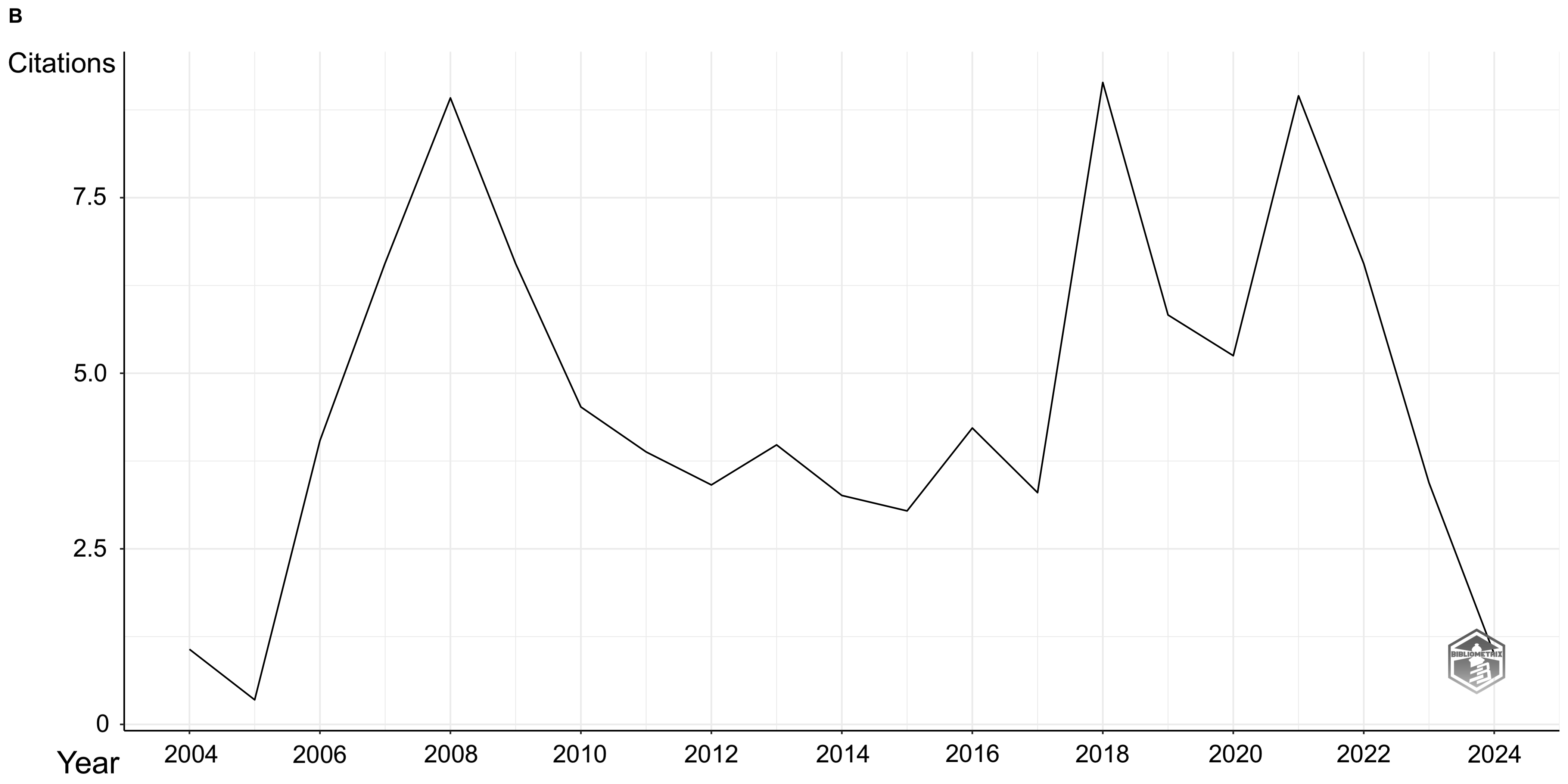}
            \caption{\textbf{Annual Scientific Production (A) and Average Citations per Year (B).} This figure depicts the yearly increase in the number of scientific publications and the average citations from 2004 to 2024.}\label{fig:annualscientificproductioncitations}
        \end{figure}

        In figure \ref{fig:annualscientificproductioncitations}, an overview of the publication trends is provided. We focus on the analysis of the annual number of publications, showcasing any significant increases or decreases in research activity over time.
    The exponential rise in publications on the EU ETS market began in 2005, corresponding with the start of phase 1. Since then, the number of articles has oscillated but consistently remained above 10 articles per year.
    An increase in the scientific production has occurred in 2021, corresponding with the start of phase 4 and the later announcement of EU ETS 2, resulting in a peak of 35 articles in 2022.
%
    \subsection{Geographic distribution}
    \label{ssec:geographic}
\begin{figure}[h]
        \centering
            \includegraphics[width=0.8\textwidth]{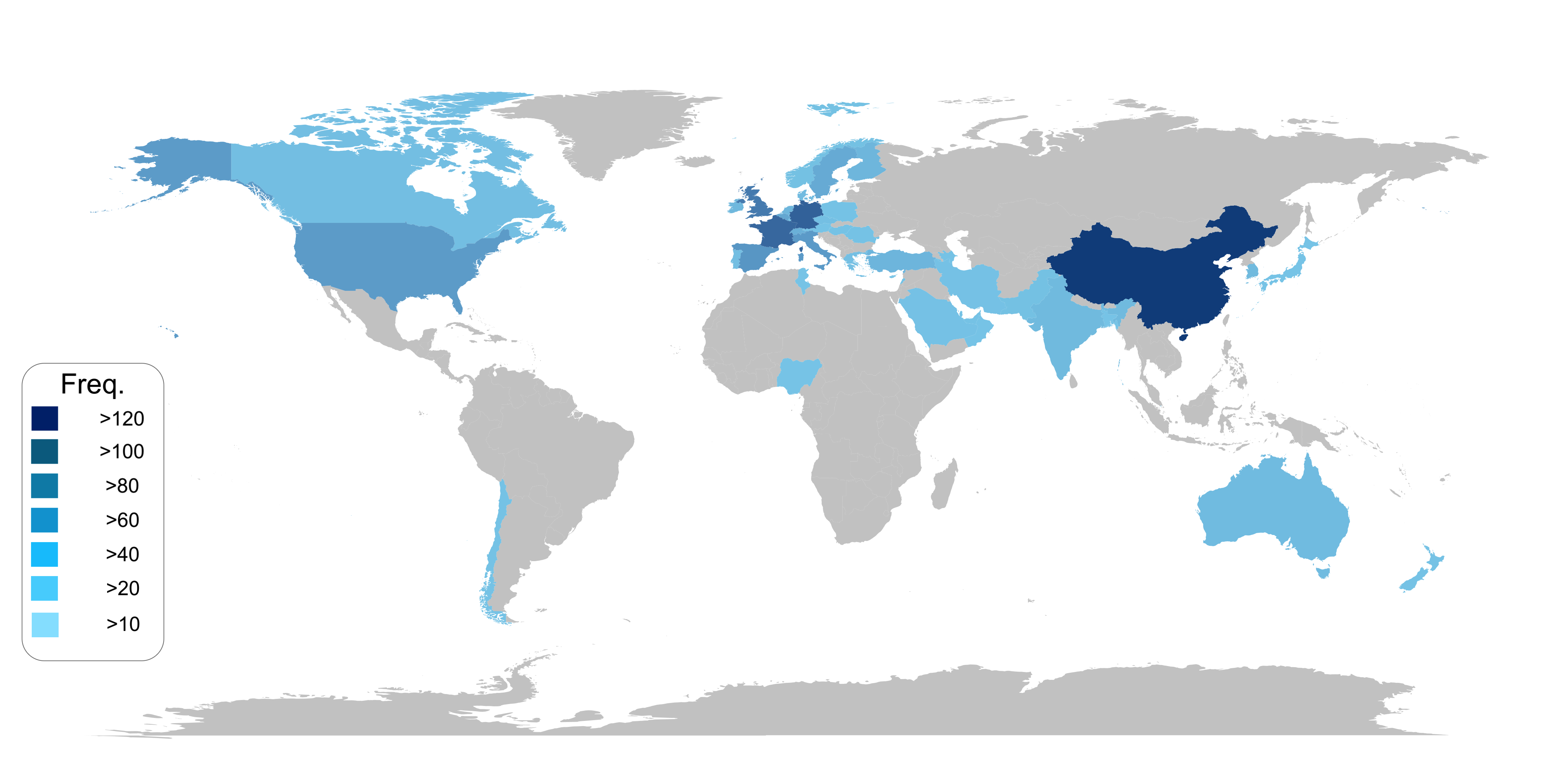}
            \caption{\textbf{Country scientific production} This plot presents the volume of scientific publications related to the EU ETS from various countries. Detailed numbers for each country can be found in table \ref{table:country_freq}.}
        \label{fig:country_freq}
        \end{figure}

        The geographic distribution analysis, presented in figure \ref{fig:country_freq}, identifies the countries that have contributed the most to the literature, based on the affiliation of the authors.
         Data on the frequency of scientific publications by region, focusing on their contribution to the global scientific community is presented in table \ref{table:country_freq}.
         China stands out as the most prolific country with a total of 168 publications, indicating a strong position in scientific output among the other countries.
         Furthermore, Germany covers the second place with 113 publications covers the second place, reflecting its significant contribution to global scientific research. Next is France with a total of  105 publications while, UK holds the fourth position with 81 publications. Hence, Spain occupies the fifth position while Italy the sixth place, with 49 and 44 publications respectively, showing important but slightly lower contributions compared to the leading nations.
        \begin{table}[h]
        \centering
        \begin{tabular}{lrrr}
        \toprule
        \textbf{Country} & \textbf{Frequency} & \textbf{Total Citations} & \textbf{Average Article Citations} \\
        \midrule
        China & 168 & 2003 & 32.80 \\
        Germany & 113 & 2476 & 55.00 \\
        France & 105 & 2150 & 53.80 \\
        UK & 81 & 1290 & 44.50 \\ 
        Spain & 49 & 932 & 44.40 \\ 
        Italy & 44 & 445 & 22.20 \\
        USA & 43 & 399 & 33.20 \\ 
        Belgium & 30 & 290 & 29.00 \\ 
        Sweden & 26 & 387 & 32.20 \\ 
        Switzerland & 20 & 553 & 55.30 \\
        Finland & 14 & 336 & 67.20 \\ 
        Netherlands & 14 & 386 & 96.50 \\
        \bottomrule
        \end{tabular}
        \caption{\textbf{Publication Frequency and Citations by Country.} This table summarizes the frequency of academic publications by country, and citations.}
        \label{table:country_freq}
        \end{table}
        In addition, USA follows with 43 publications, Belgium, Sweden, and Switzerland have 30, 26, and 20 publications respectively. Finally, Finland and Netherlands show 14 publications each.
        Furthermore, table \ref{table:country_freq} provides a ranking of countries based on their research citation metrics, reflecting both the total number of citations and the average citations per research output.
        The analysis indicates that Germany leads with the highest total citations and a high average of 55.00 citations per paper.
        France and China follow, with France having slightly more total citations but a higher average citation rate compared to China. The Netherlands stands out with the highest average citation rate, 96.50, despite having fewer total citations, 386.
%
    \subsection{Venue analysis}
    \label{ssec:venue}
        \begin{table}[h]
        \centering
        \begin{tabular}{lc}
        \toprule
        \textbf{Venues} & \textbf{Articles} \\ 
        \midrule
        Energy Policy & 64 \\
        Energy Economics & 61 \\
        Environmental and Revenue Economics & 17 \\
        Journal of Environmental Economics and Management & 15 \\
        Environmental Science and Pollution Research & 11 \\
        International Journal of Global Energy Issues & 11 \\
        Energy Journal & 10 \\
        Finance Research Letters & 9 \\
        Journal of Cleaner Production & 8 \\
        Revenue and Energy Economics & 8 \\
        \bottomrule
        \end{tabular}
        \caption{\textbf{Most relevant venues per articles.} This table shows the most prolific sources by number of published articles.}
        \label{table:relevant_venues}
        \end{table}
        In table \ref{table:relevant_venues}, a list of the top ten venues extracted from our dataset, with the respective number of articles on EU ETS published, is presented.
        The leading Journal is Energy Policy with a total of 64 articles, followed closely by Energy Economics with 61 articles. These two journals stand out significantly in terms of article volume, highlighting their important role in disseminating research on the EU ETS and energy markets in general.
        However, the remaining venues show a significant drop in article numbers, with the third-ranked journal, Environmental and Revenue Economics, publishing 17 articles.
        The other journals, while still contributing valuable research, appear to have less emphasis on EU ETS economics and finance topics, as evidenced by their lower article counts.
        \begin{figure}[h]
        \centering
            \includegraphics[width=0.8\textwidth]{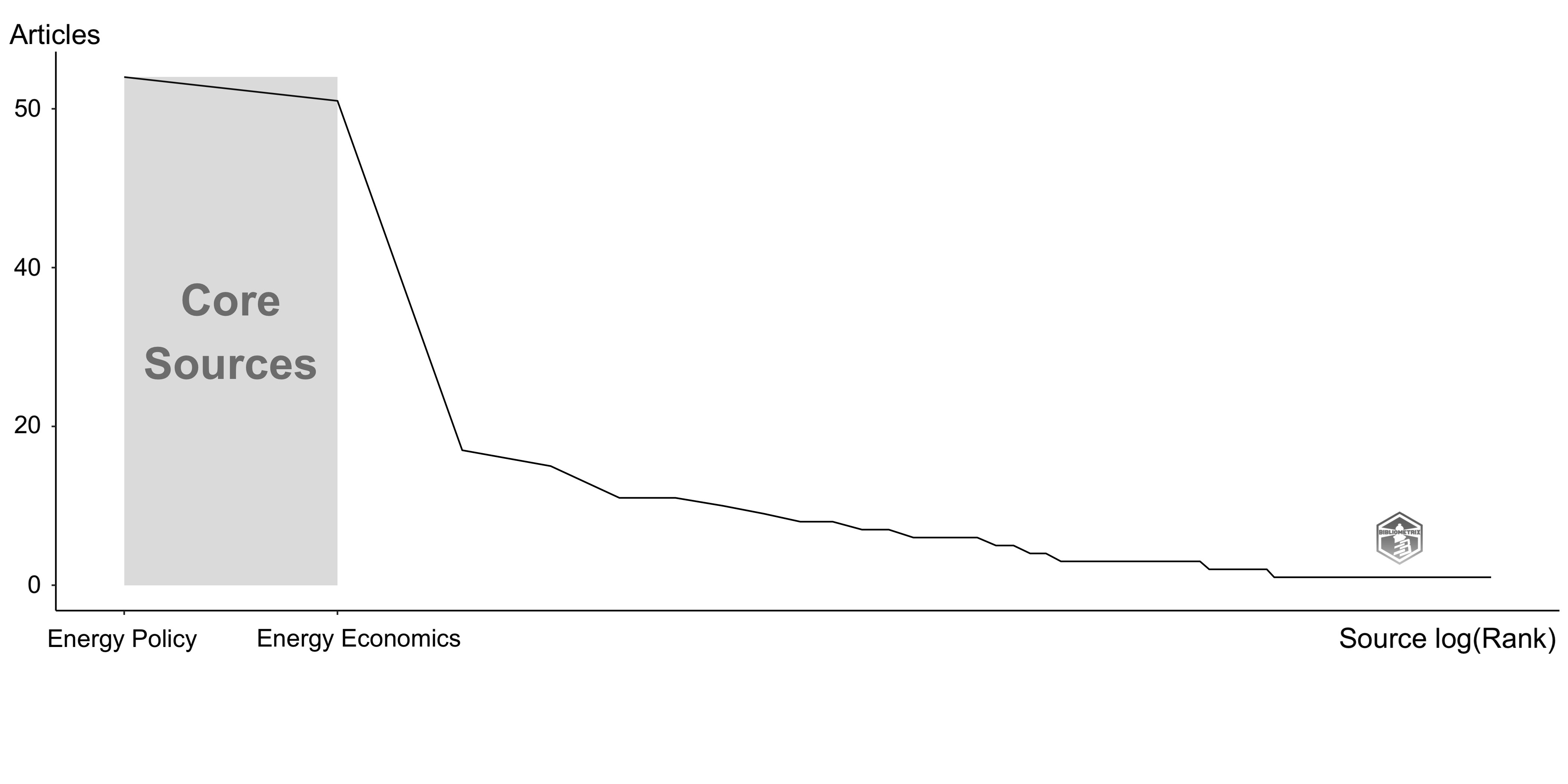}
            \caption{\textbf{Core Sources by Bradford's Law.} This plot illustrates the Bradford's Law metric, which describes the distribution of research articles across a set of sources.}
        \label{fig:BradfordLaw}
        \end{figure}
        The Bradford's Law is a bibliometric metric that maps how scientific literature is distributed across venues. Formulated originally by \cite{bradford1934sources}, the metric suggests that in any given field of study, articles are not uniformly distributed among all journals. Instead, a small group of journals will contain the majority of articles on a specific topic, while a larger number of journals will each contribute a smaller number of articles.
        In this context, there is a clear predominance of the first two venues, Energy Policy and Energy Economics, which lead the article count, reflecting their significant impact and centrality in the field compared to other journals.
%
    \subsection{Authors analysis}
    \label{ssec:authors}
        \begin{figure}[h]
        \centering
            \includegraphics[width=0.8\textwidth]{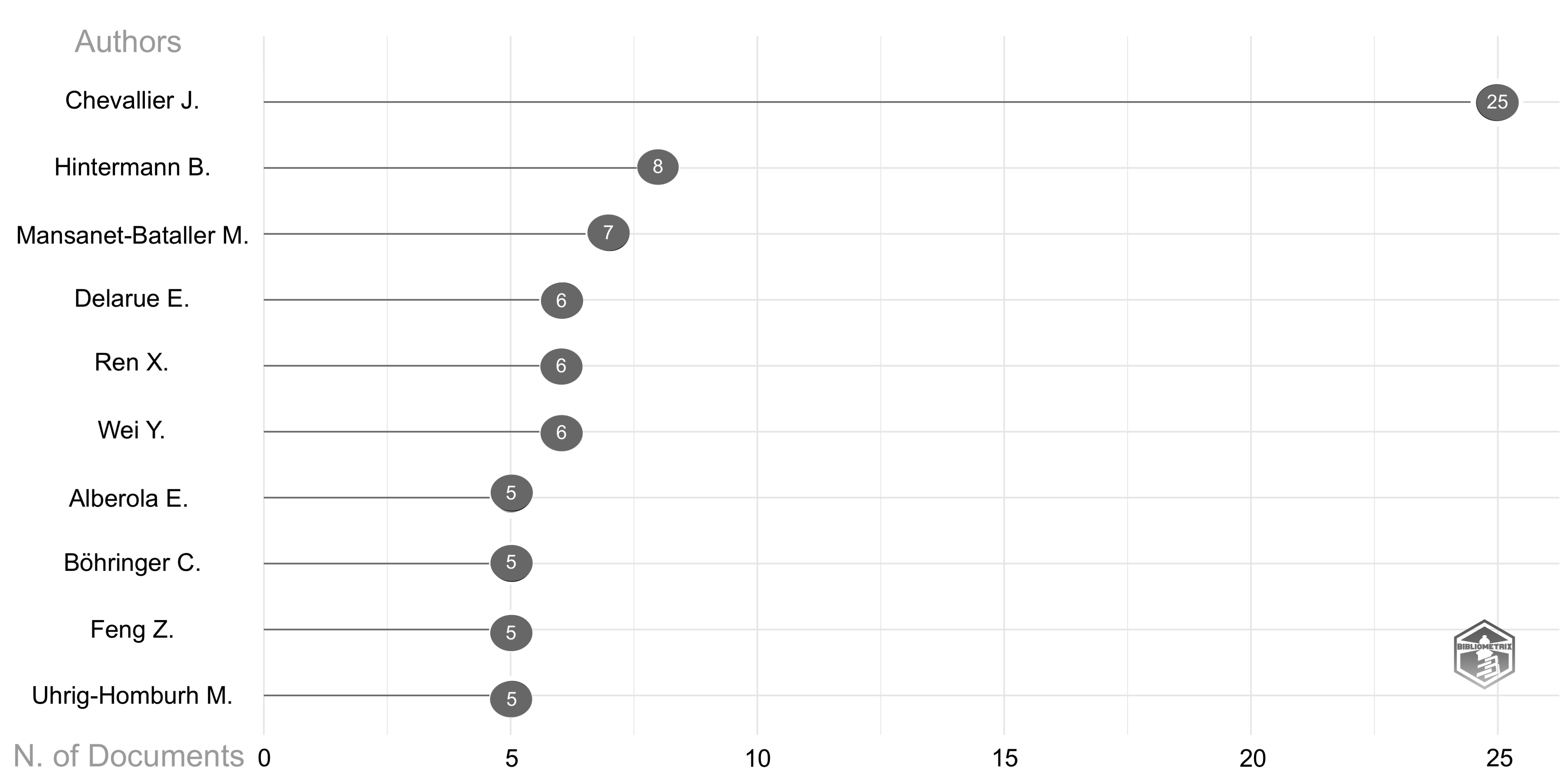}
            \caption{\textbf{Most relevant Authors.} This plot highlights the prolific authors in the field, demonstrating their contributions through the number of publications and the impact of their work, measured by citation counts.}
        \label{fig:RelevantAuthors}
        \end{figure}
        The top 10 locally cited authors are presented in figure \ref{fig:RelevantAuthors}.
        The leading author, Chevallier J., has published 25 articles, with a fractionalized count of 13.90, demonstrating their strong influence and active involvement in European Exchange Trading System research.
        Moreover, Hintermann B. and Mansanet-Bataller M. are also notable contributors with 8 and 7 articles, with a fractionalized counts of 5.67 and 2.58, respectively.
        Furthermore, other authors, such as Delarue E., Ren X., and Wei Y., each have 6 articles, with their fractionalized counts ranging from 1.43 to 2.00.
        The remaining authors, including Alberola E., Böhlinger C., Feng Z., and Uhrig-Homburg M., each have 5 articles. Their fractionalized counts range from 1.57 to 2.57, showcasing their relevant but more modest influence in the field compared to the leading authors.
        \begin{figure}[h]
        \centering
            \includegraphics[width=0.8\textwidth]{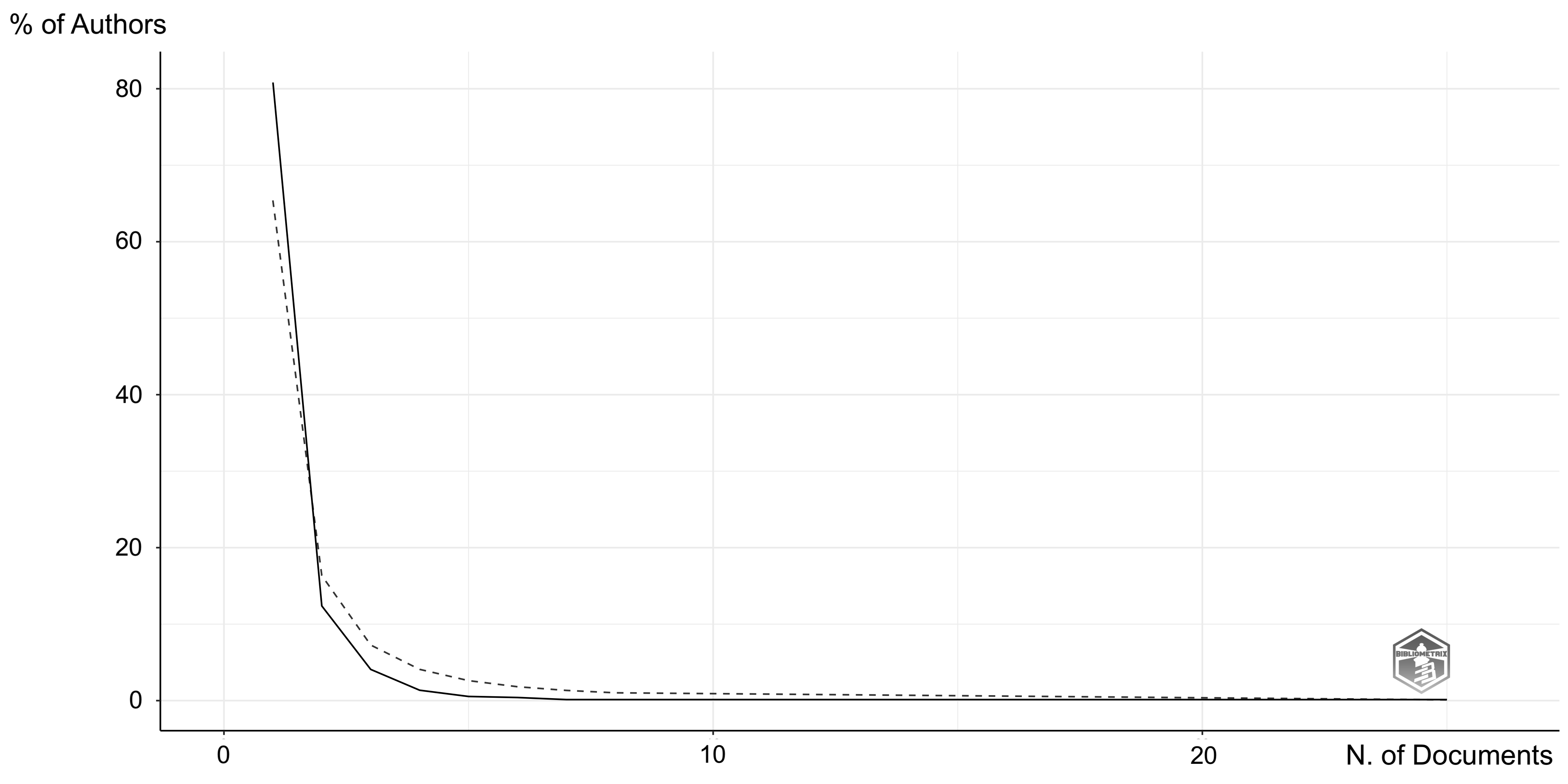}
            \caption{\textbf{Authors productivity through Lotka's Law.} This figures presents the Lotka's Law metric, which describes the distribution of scientific productivity among authors.}
        \label{fig:LotkaLaw}
        \end{figure}
        Lotka's Law by \cite{lotka}, is presented in figure \ref{fig:LotkaLaw}. This indicator shows author productivity in a scientific field, which states that the number of authors publishing a certain number of papers is inversely proportional to the square of that number. This means that a small number of authors are highly productive, while most authors publish fewer works, resulting in a power-law distribution. Thus, scientific output is concentrated among a few prolific authors, like Chevallier J., while the majority contribute less.
        \begin{figure}[h]
        \centering
            \includegraphics[width=0.8\textwidth]{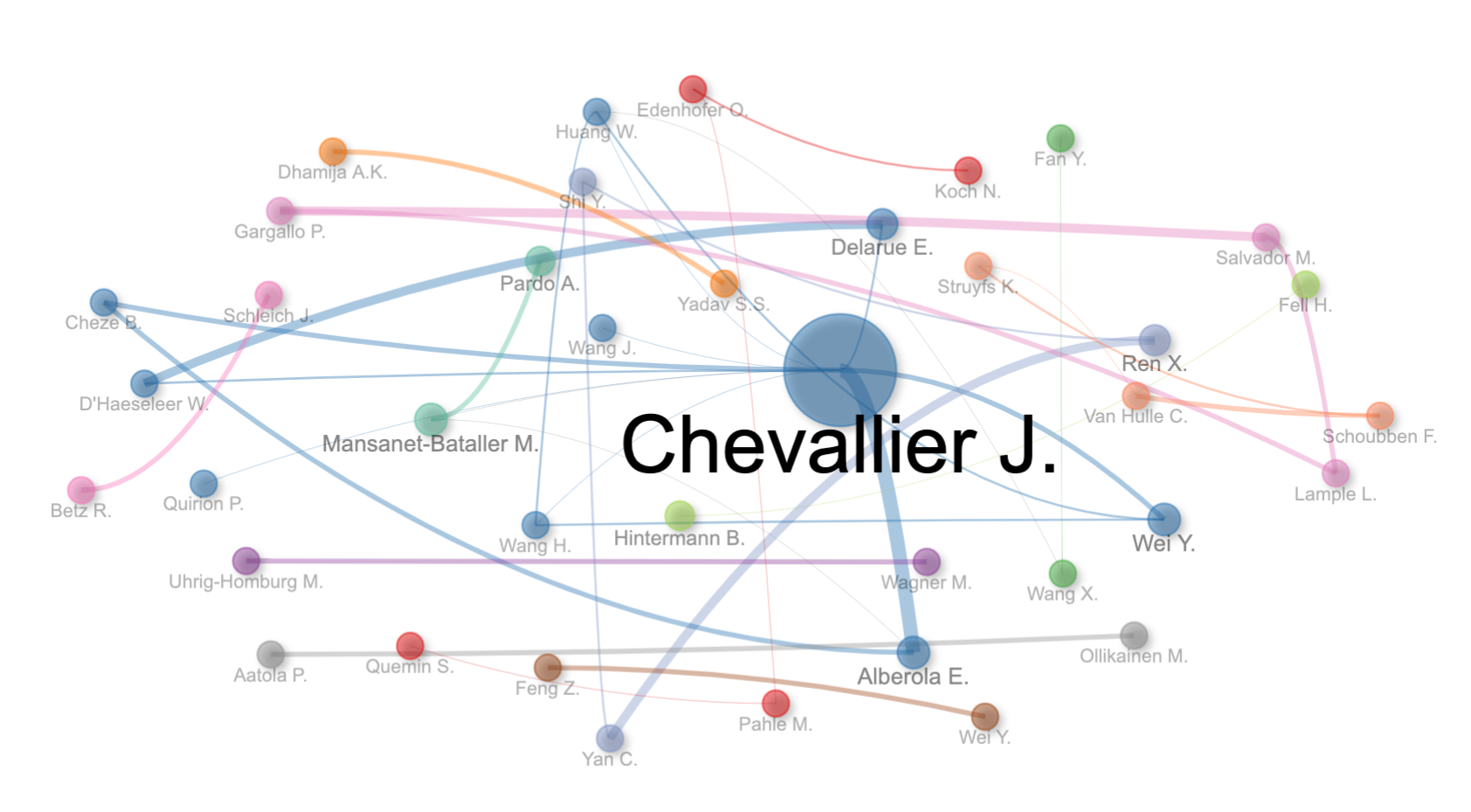}
            \caption{\textbf{Authors collaboration network.} This network maps the research relationships between authors, showing the strong connections built through co-authorship and research groups among scholars. Each node represents an author, and the lines between them show their collaborations. The thickness of the lines shows how often they worked together, highlighting strong partnerships and the wide network of ideas in the field.}
        \label{fig:AuthorsNetwork}
        \end{figure}
        Figure \ref{fig:AuthorsNetwork} shows a network analysis of author collaborations, highlighting key metrics such as cluster membership, betweenness centrality, closeness centrality, and PageRank for each author.
        In Cluster 1, Edenhofer O. and Pahle M. have betweenness values of 2, indicating they serve as key intermediaries in the network. Their closeness values of 0.25 reflecting strong connectivity within the cluster.
        In Cluster 2, Chevallier J. stands out with a high betweenness of 62, indicating a key role in connecting different parts of the network, although their closeness of 0.0588 suggests fewer direct connections. With the highest PageRank in the dataset at 0.0809, Chevallier J. holds significant influence. Wei Y. and Huang W. have betweenness values of 22 and 0, respectively, with closeness values of 0.04 and 0.0435, showing moderate connectivity and influence. Alberola E. has a betweenness of 1 and a closeness of 0.0385, reflecting a moderate role in network intermediation.
%
\section{Document analysis} 
\label{sec:document_analysis}
%
    \subsection{Most cited documents}
    \label{ssec:cited_documents}
        \begin{figure}[h]
        \centering
            \includegraphics[width=0.8\textwidth]{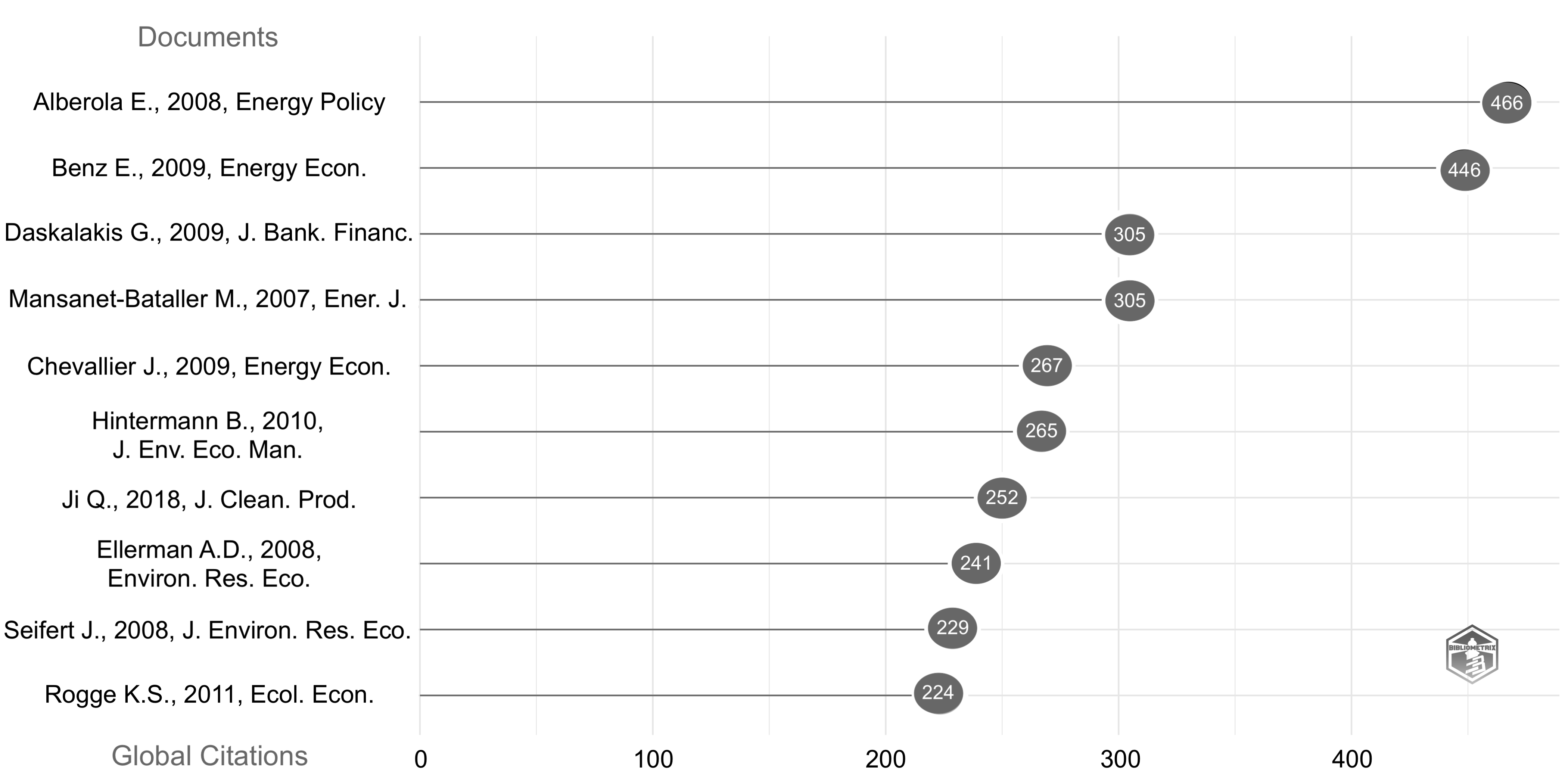}
            \caption{\textbf{Most global cited documents.} Overview of the top 10 most globally cited documents. The plot provides the number of global citations for each paper.}
        \label{fig:GlobalCitedDocuments}
        \end{figure}
    According to figure \ref{fig:GlobalCitedDocuments}, the most cited paper is \cite{Alberola2008787} in Energy Policy, with 466 Total Citations (TC) and a high TC per year of 27.41.
    Closely following is \cite{Benz20094} in Energy Economics, with 446 citations, the highest TC per year of 27.88, and a normalized TC of 4.25, showing it's also highly influential.
    The paper by \cite{Creti2012327} in Energy Economics has the highest normalized TC of 4.62, showing it is highly influential when considering the field and publication year. Also, \cite{Ji2018972} in the Journal of Cleaner Production has a notable citation rate of 36.00 per year and a high normalized citation count of 3.94.
    Moreover, other research articles from \cite{Mansanet-Bataller200773} and \cite{Chevallier2009614} show consistent citations over the years, with TC per year values of 16.94 and 16.69, respectively, indicating steady recognition in the field. The citations come from a range of journals, including Energy Policy, Energy Economics, Journal of Environmental Economics and Management, and Ecological Economics.
%
    \subsection{Word analysis}
    \label{ssec:word_analysis}
\begin{figure}[h]
\centering
    \begin{minipage}{0.48\textwidth}
        \centering
        \includegraphics[width=0.8\textwidth]{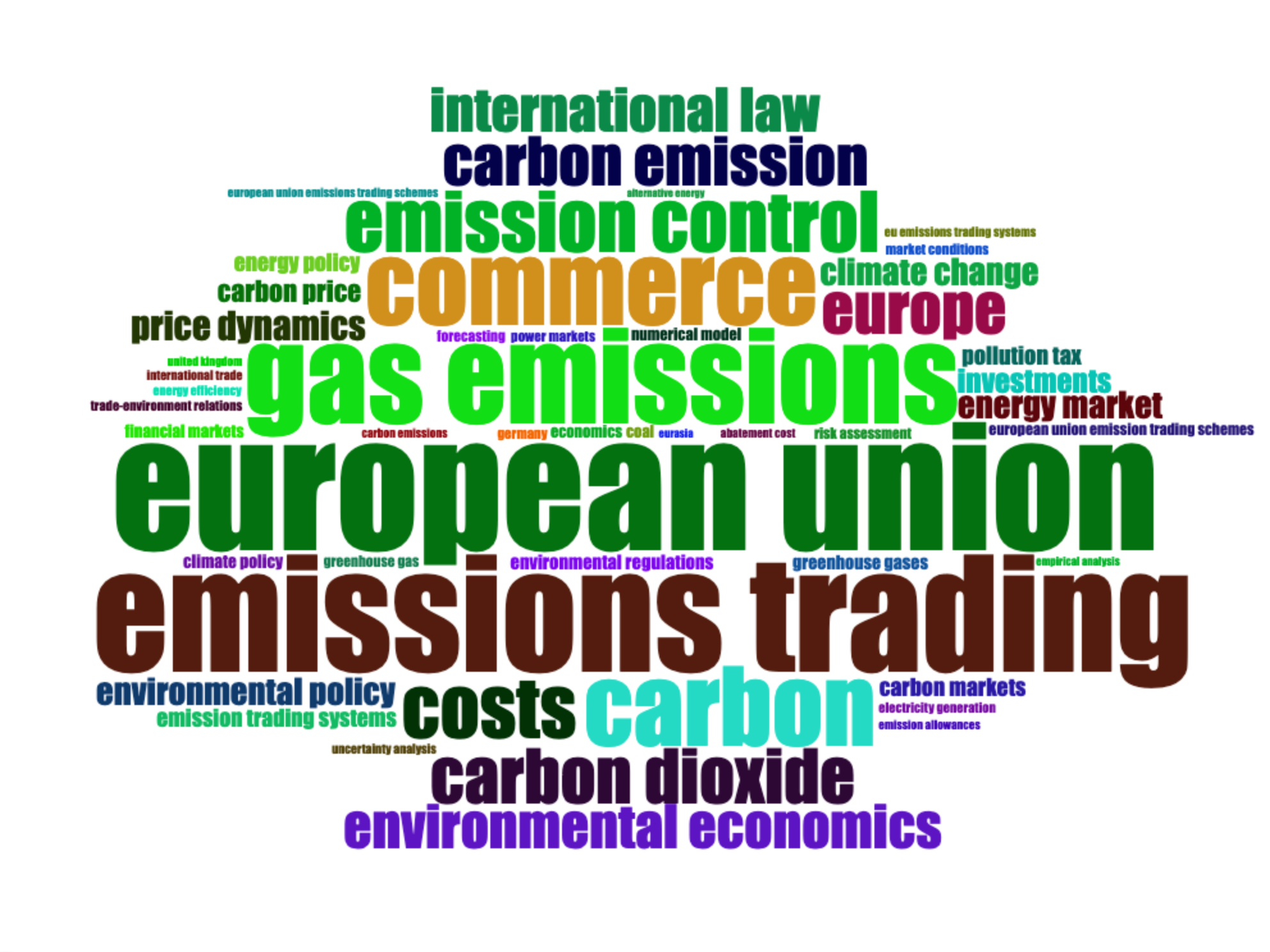}
    \end{minipage}%
    \hfill
    \begin{minipage}{0.48\textwidth}
        \centering
        \resizebox{\textwidth}{!}{%
        \begin{tabular}{llrl}
        \toprule
        \textbf{Terms} & \textbf{Freq.} & \textbf{Terms} & \textbf{Freq.} \\
        \midrule
        EU ETS & 141 & Carbon Futures & 9 \\
        Emission Trading & 40 & Market Stability Reserve & 9\\
        Carbon Price & 28 & Carbon Trading & 8 \\
        Climate Change & 17 & European Union & 8 \\
        Carbon Market & 16 & $\text{CO}_{2}$ Emissions & 7 \\
        Climate Policy & 15 & Renewable Energy & 7 \\
        EUA & 14 & Volatility & 7 \\
        Carbon Pricing & 13 & Competitiveness & 6 \\
        Carbon Leakage & 12 & Emission Trading Scheme & 6 \\
        Cap-and-Trade & 11 & Energy Markets & 6 \\
        Energy Prices & 10 & & \\
        \bottomrule
        \end{tabular}
        }
    \end{minipage}
    \caption{\textbf{Word cloud and terms frequency.} The left side displays a word cloud highlighting key themes, while the right side presents the frequency of specific terms found in the textual data.}
    \label{fig:combined}
\end{figure}
\begin{figure}[h]
        \centering
            \includegraphics[width=0.8\textwidth]{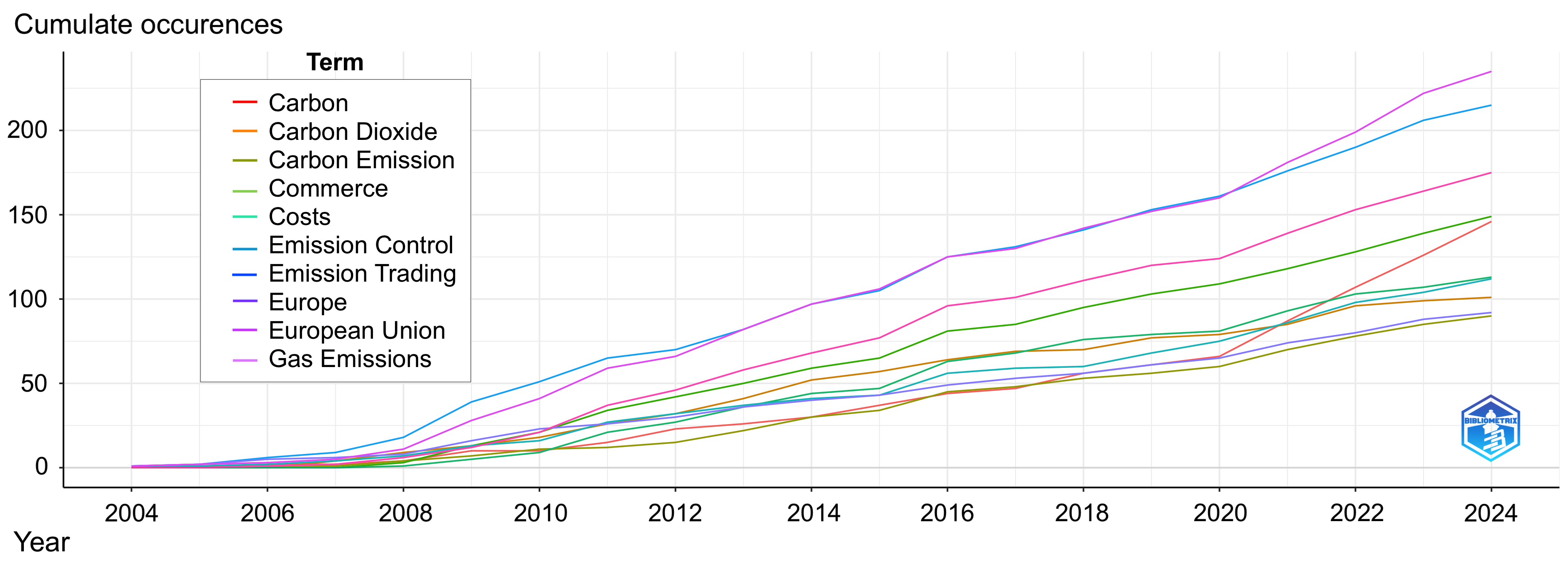}
            \caption{\textbf{Word frequency over time.} This plot illustrates the increasing frequency of key terms related to emissions, carbon, and environmental policy over a two-decade period, highlighting the growing emphasis on these themes.}
        \label{fig:wordfrequencyovertime}
        \end{figure}

        The analysis in Figure \ref{fig:combined} highlights the most common terms in the dataset, giving insight into key themes in EU ETS research. Terms like European Union, emissions trading, and gas emissions show a focus on regulations and environmental concerns. Frequent mentions of carbon, costs, and emission control point to an emphasis on the economic and practical aspects of managing emissions. Additionally, terms like environmental economics, international law, and price dynamics suggest interdisciplinary research connecting economics, law, and market behavior to climate issues.
        Figure \ref{fig:wordfrequencyovertime} presents the frequency of various terms related to emissions and environmental policies over the years of investigation. The term \textit{European Union} has shown a steady increase in frequency, starting from 1 occurrence in 2004 and reaching 235 occurrences in 2024.
        Similarly, \textit{Emission trading} has also increased from 1 occurrence in 2004 to 215 occurrences in 2024. \textit{Gas emissions} and \textit{Commerce} have followed a similar trend, starting from very few or no occurrences in the early years and growing to 175 and 149 occurrences, respectively, by 2024.
        The term \textit{Carbon} has increased from no occurrences in 2004 to 146 occurrences in 2024. \textit{Costs}, \textit{Emission control}, and \textit{Carbon dioxide} have also shown significant increases over the years, reaching 113, 112, and 101 occurrences, respectively, in 2024.
        The term \textit{Europe} has seen a steady rise from 1 occurrence in 2004 to 92 in 2024, while \textit{Carbon emission} has increased from no occurrences in 2004 to 90 in 2024.
%
\section{Conceptual structure}
\label{sec:conceptualstructure}
%
    \subsection{Network analysis}
    \label{ssec:Networkapproach}
        \begin{figure}[h]
        \centering
            \includegraphics[width=0.8\textwidth]{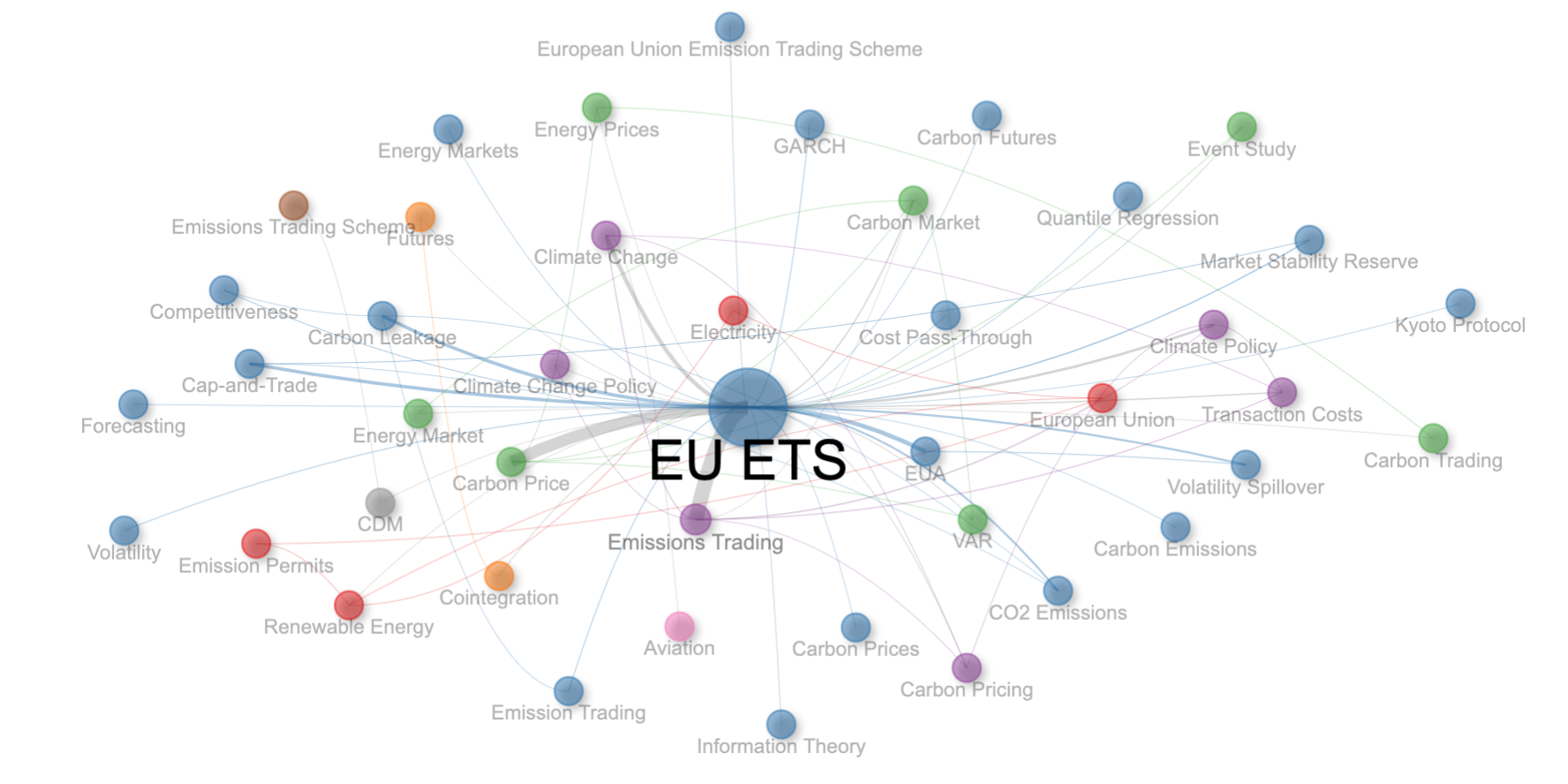}
            \caption{\textbf{Co-occurrence network analysis.} This co-occurrence network maps the landscape of author keywords across the academic literature examined. The network plot highlights how frequently keywords co-occur, illustrating the central themes and their interrelationships, providing insights into evolving research trends and thematic clusters within the field.}
        \label{fig:Cooccurrencenetwork_author}
        \end{figure}
     Figure \ref{fig:Cooccurrencenetwork_author} presents the co-occurrence network analysis based on author keywords. Different groups of keywords that usually are connected in scientific literature can be found, with a strong predominance of the blue group. Two measure are discussed, the PageRank value that measures the influence of author keywords based on their connections, highlighting their centrality and importance within the network and between which quantifies how often a node acts as a bridge between other nodes, indicating its role in connecting different parts of the network.
     In these networks, certain keywords, such as \textit{EU ETS} and \textit{Kyoto Protocol}, exhibit high betweenness values, highlighting their leading role as intermediaries that connect disparate areas of the network.
     Specifically, \textit{EU ETS} stands out with a remarkably high betweenness value of 3437.60, positioning it as a central hub with significant influence over the flow of information. This centrality is further reinforced by its substantial PageRank value of 0.273, which demonstrates its significant role in organizing the network.
    In contrast, nodes like \textit{carbon prices} and \textit{Energy markets} show lower betweenness and PageRank values, meaning they are less central but still important to the network. This suggests a network with dominant themes, while other topics support the overall structure.
%
    \subsection{Thematic analysis}
    \label{ssec:thematicAnalysis}
        \begin{figure}[h]
        \centering
            \includegraphics[width=0.8\textwidth]{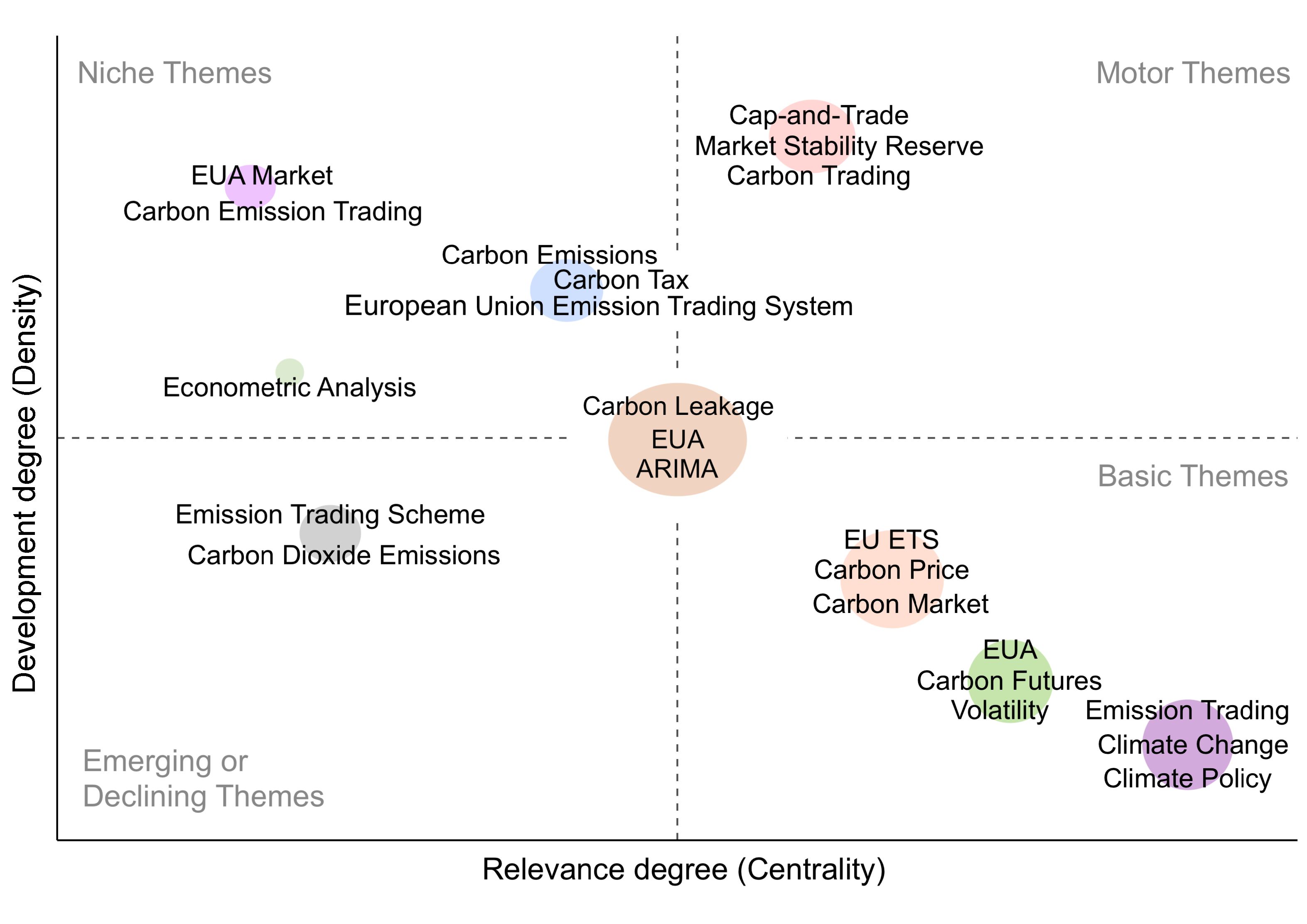}
            \caption{\textbf{Thematic map analysis.} The thematic map highlights the distribution and centrality of key terms across different clusters. The research areas are classified into four main types of research themes: niche themes, motor themes, basic themes and emerging or declining themes.}
        \label{fig:thematicmap}
        \end{figure}
        Figure \ref{fig:thematicmap} provides a clear view of the main themes and their importance within the co-occurrence network based on author keywords.
        In Cluster 1 (\textit{cap-and-trade}), key themes like \textit{Cap-and-Trade}, \textit{Carbon trading}, and \textit{Emissions Trading Scheme} are central to discussions on carbon markets and regulations. The high betweenness centrality of \textit{Cap-and-Trade} (190.19) highlights its key role in connecting various ideas in this area. However, terms like \textit{EU ETS reform} have lower centrality, suggesting they play a more specialized or less influential role within this cluster.
        Cluster 1 centers on cap-and-trade systems, with key terms acting as important connectors. Cluster 2 focuses on carbon emissions, carbon tax, and electricity markets, highlighting the economic and policy aspects. Terms like \textit{carbon emissions} and \textit{Carbon tax} play key roles in connecting ideas about carbon pricing and market impacts.
        Cluster 2 also includes specialized terms like \textit{Cost pass-through} and \textit{Market power}, which add depth to the discussion on carbon economics. Cluster 3 centers on the European Union Allowance (EUA) and related topics, with terms like \textit{EUA}, \textit{Carbon futures}, and \textit{Volatility} playing central roles in the network.
        The term \textit{EUA} has a high Betweenness Centrality of 43.37, highlighting its key role in EU carbon market discussions. Similarly, \textit{Carbon futures} and \textit{Volatility} are important, as their centrality shows their relevance to market fluctuations and financial aspects of carbon trading.
        Overall, the thematic map analysis shows the varying influence of terms across clusters. Cluster 1 focuses on cap-and-trade system policy, Cluster 2 on the economic aspects of carbon emissions, and Cluster 3 on European carbon market mechanisms.
%
\section{Discussion and Future research}
\label{sec:discussion}
    Over the past two decades, there has been a noticeable increase in both the frequency and impact of certain key terms and topics, which have played a central role in shaping the discussion on emissions trading. A strong focus on market mechanisms and regulations in the European Union reflects a shared interest among researchers and policymakers in using carbon pricing as a key tool to combat climate change. Foundational works by \cite{Alberola2008787} and \cite{Benz20094} have emerged as some of the most cited contributions, proving crucial advancements the field, especially regarding market dynamics and policy impacts. Their influence is evident from their high citation metrics, highlighting their role in shaping the academic and policy understanding of the European carbon market.
    Furthermore, the word frequency analysis emphasizes a growing focus on terms such as \textit{European Union}, \textit{Emissions Trading}, and \textit{Carbon}, which reflects the increasing importance of carbon pricing mechanisms and the carbon market within the broader context of climate policy. This trend illustrates the centralization of climate policy discussions around carbon pricing as an economically viable solution to reducing emissions on a large scale. The repeated appearance of these terms over time suggests a stable, if not growing, alignment between economic and environmental policy-making communities on the potential of emissions trading systems.
    In parallel, there is an increased research emphasis on specific market behaviors and their implications for policy, as indicated by the rising frequency of terms such as \textit{Carbon Price} and \textit{Carbon Market}. This interest signals an intensifying examination of how carbon pricing affects not only emissions levels but also economic factors, such as sectoral competitiveness, cost-effectiveness, and overall market behavior under regulatory constraints. The evolving focus on pricing strategies and economic effects reveals the complexity of balancing environmental goals with economic stability, a recurring theme that has received significant attention in recent literature.
    Moreover, trend analysis uncovered a gradual shift in research focus toward the financial aspects of the EU ETS. Topics such as \textit{Carbon Leakage}, \textit{Energy Prices}, and \textit{Carbon Futures} have seen increasing attention, indicating that scholars are now more engaged with the financial and secondary effects of carbon trading, beyond just the environmental or policy themes. This shift reflects a nuanced understanding of the EU ETS, which now extends beyond emissions control to cover complex financial dynamics, such as market volatility, investor behavior, and risk assessment in carbon trading. These financial factors show the EU ETS’s growing role in the economy, revealing emissions trading as both a regulatory tool and a financial instrument influenced by market trends and economic shifts.
    The network and thematic map analyses provide a clearer view of the conceptual structure within the EU ETS literature. Central themes like \textit{Carbon Leakage} and \textit{EUA} act as critical nodes that connect diverse research themes, highlighting their importance in shaping the research activity. Their high Betweenness Centrality and PageRank values emphasize their impact in linking different areas of research. This connectivity reflects a common interest in understanding indirect consequences of emissions trading, such as shifts in production patterns, competitiveness, and trade imbalances. In contrast, terms like \textit{Carbon Prices} and \textit{Energy Markets}, while influential, play more supportive roles in the network. This suggests potential growth opportunities for future research that could deepen our understanding of these secondary yet interconnected factors.
    Despite these important contributions, several research gaps remain unaddressed.
    There is a noticeable gap in the literature regarding the use of non-parametric methodologies, particularly in the area of feature selection and forecasting models. This gap suggests that traditional, parametric approaches may be limited in capturing the nonlinear, often complex relationships inherent in emissions trading data. For example, by employing non-parametric quantile regression based on copulas, \cite{SEGURA20181111} captures non-linear relationships and dependencies among variables while controlling for firm characteristics to address omitted variable biases. In particular, non-parametric methods may better model and capture extreme market conditions, which have become more frequent due to recent economic events and policy shifts. Another non-parametric approach is introduced by \cite{Salvagnin2024}, who examines how the COVID-19 pandemic and energy crises have altered market dynamics, making EU ETS prices more sensitive to energy factors. The study also proposes a new non-parametric method for feature selection and forecasting, using the information content of financial variables. Additionally, \cite{Lan2024} introduces a novel secondary decomposition-integration framework, validating the effectiveness of nonlinear integration and optimization methods to enhance model performance. Although non-parametric approaches offer flexible alternatives to parametric methods, they are still relatively unexplored in emissions trading contexts.
    Emissions trading involves complex and often incomplete datasets, and methodologies for effective imputation and aggregation are underdeveloped in the EU ETS literature. For instance, \cite{su15086394} imputed missing values to include high-emitting installations that are likely to lose free allocations due to policy changes, leading to more accurate regulatory impact assessments. This approach prevents underestimation of carbon leakage dynamics \cite{Biancalani2024}. By refining imputation methods, future research could produce more reliable datasets, allowing for more precise assessments of market trends and policy impacts. Enhanced data quality would contribute to the robustness of both theoretical models and practical policy recommendations.
    Another possible research gap lies in examining how exchange rates interact with EU ETS dynamics, which could be particularly insightful for understanding cross-border trade effects and market competitiveness. In a recent study, \cite{MORADIANDAGHIGH2023139043} examines the influence of exchange rates on carbon prices, identifying the Euro index as a significant driver, especially during the Brexit crisis. Currency fluctuations can signal broader macroeconomic conditions that may affect carbon market stability. Future research could investigate how global economic shifts impact the carbon market, offering insights into potential vulnerabilities and providing guidance for investors and policymakers on mitigating currency-related risks.
    Macroeconomic factors like inflation, GDP growth, and interest rates may affect carbon allowance pricing and trading. \cite{BOLAT2023106879} highlights the significant influence of these conditions on carbon price dynamics, underscoring the need to integrate these variables when evaluating the EU ETS’s effectiveness in meeting climate targets. By understanding these macroeconomic impacts, researchers could help clarify how emissions trading performs under various economic cycles, potentially leading to more resilient policy frameworks that adapt to economic fluctuations.
    Another remarkable gap lies in the exploration of causal relationships among the EU ETS and other financial variables. In a recent work, \cite{CANDELON2023103878} applied Granger causality tests in the frequency domain, distinguishing short- and long-term effects of climate policies. Financial instruments such as derivatives, equities, and bonds could be influenced by, or have an influence on, the dynamics of the EU ETS. \cite{ren2023} found that causality between crude oil and stock markets varies by equity index, remaining robust under normal conditions but weakening in extreme market scenarios. As emissions trading markets mature, understanding these causal relationships could reveal important insights into how external financial shocks and policy changes influence the ETS, guiding future regulatory and investment strategies.
    In summary, these gaps and trends highlight key areas for future research to improve our understanding of the EU ETS. Addressing the methods, data, economic factors, and causal relationships in emissions trading is essential for understanding how the EU ETS can better support carbon reduction and climate policy goals. By focusing on these areas, we can enhance the effectiveness and adaptability of emissions trading systems in response to changing global conditions.
%
\section{Conclusions}
\label{sec:conclusions}
    The EU ETS, is a key part of Europe's climate policy, using a cap-and-trade system to control greenhouse gas emissions. The current phase (2021-2030) continues to tighten emission limits and cover more sectors to keep the EU on track with its climate targets.
    Our bibliometric review, covering 367 papers from 2004 to 2024 in the Scopus database, gives a detailed look at academic research on the EU ETS. Using tools like citation analysis, co-authorship networks, and keyword analysis, we highlight publication trends, collaboration, and changing research topics in this field.
    The descriptive analysis reveals an annual growth rate of 12.99\%, with notable peaks linked to key policy changes and regulatory updates. Countries like China, Germany, and France have emerged as major contributors, reflecting the global interest in emissions trading systems. Germany leads in total citations, while the Netherlands has the highest average citation rates, signifying the impact and influence of research from these countries. Leading venues, such as \textit{Energy Policy} and \textit{Energy Economics}, play fundamental roles in disseminating research.
    The review highlights a growing research focus on topics such as carbon pricing, market volatility, and the financial aspects of carbon trading, which reflect the increasing complexity of carbon markets. Key terms like \textit{Carbon price}, and \textit{Carbon market} have gained prominence over time, pointing to an intensified interest in pricing strategies, market behavior, and the broader economic implications of emissions trading systems.
    The network and thematic map analyses show the main ideas in EU ETS research, with key topics like \textit{Carbon leakage}, \textit{EUA}, and \textit{ARIMA} serving as central points linking different research areas. The main themes found cap-and-trade systems, economic aspects of carbon emissions, and financial topics related to the European Union Allowance highlight how the research is evolving, with more focus on the financial and side effects of emissions trading.   
    Despite these advances, several research gaps remain. There is not enough focus on non-parametric methods, especially in feature selection and forecasting. Methods for implementing mixed-frequency data are also lacking, which are important for stronger analysis, incorporating different economic factors. Another gap is the lack of research on how exchange rates affect the EU ETS market.
    In addition, the effects of macroeconomic factors like inflation, interest rates, and GDP growth on the EU ETS have not been thoroughly studied, offering an important area for future research. There's also a need for more causal analysis of how the EU ETS interacts with other financial variables.
    In conclusion, this bibliometric review highlights the evolving and increasingly complex research on the EU ETS. While there has been significant progress in understanding market dynamics, carbon pricing, and policy impacts, some gaps remain, particularly in non-parametric methods, macroeconomic factors, and financial connections. Addressing these gaps will be important for further advancing the research and supporting future policies aimed at improving the effectiveness of emissions trading systems in addressing climate change.
%
\section*{Statements and Declarations}
\label{DeclarationOfInterest}
    \noindent
    The author report there are no competing interests to declare.
%
\section*{Funding}
\label{sec:funding}
\noindent
    The research of Cristiano Salvagnin is supported by the Centre for the Analysis and Measurement of Global Risks (CAM-Risk) Project - Financial Oversight and Risk-Tailored Understanding for New Evaluation.
%
\section*{Replicating and supplementary materials}
\label{sec:ReplicatingSupplementaryMaterials}
    \noindent
    Replicating and supplementary materials can be found at the following GitHub repository: \href{https://github.com/SaveChris/EUA_BilbliometricReview}{EUA Bibliometric Review}
%
\section*{Acknowledgments}
\label{sec:Acknowledgments}
    \noindent
     We extend our sincere thanks to Aldo Glielmo (Bank of Italy), Maria Elena De Giuli (University of Pavia), and Antonietta Mira (Università della Svizzera Italiana) for their careful reading and the invaluable insights they shared.
%
\bibliographystyle{abbrv} 
\bibliography{Refs}



\end{document}